\documentclass[fleqn,10pt]{wlscirep}
\usepackage[utf8]{inputenc}
\usepackage[T1]{fontenc}
\title{Scanning Thermal Microscopy in Air and Vacuum: A Comparison}

\author[1,*]{Jabez J. McClelland}
\author[1]{Evgheni Strelcov}
\author[2]{Ami Chand}
\affil[1]{Nanoscale Device Characterization Division, National Institute of Standards and Technology, Gaithersburg, MD 20899 USA}
\affil[2]{Applied NanoStructures, Inc., 415 Clyde Avenue, Suite 102, Mountain View, CA 94043 USA}
\affil[*]{jabez.mcclelland@nist.gov}
\DeclareMathOperator\erf{erf}

\begin{abstract}
We present measurements comparing scanning thermal microscopy in air and vacuum. Signal levels are compared and resolution is probed by scanning over the edge of a nanofabricated Ag square embedded in SiO$_{2}$.  Signals measured in air were seen to be 2.5 to 40 times larger than in vacuum. Furthermore, the air signals were stable while the vacuum signals varied significantly.  Edge widths measured in air were approximately 39~\% larger than those measured in vacuum.  Our observations are consistent with the air measurements experiencing heat transfer from the surrounding sample via conduction and convection as well as the formation of a water-related meniscus at the tip-sample junction.  These results contribute to the understanding of the complex heat exchange effects that can occur in scanning thermal microscopy when it is conducted in an ambient atmosphere.            

\end{abstract}
\begin{document}

\flushbottom
\maketitle

\thispagestyle{empty}
\section*{Introduction}

\noindent Scanning thermal microscopy (SThM) has emerged in recent years as a useful tool for the study of nanoscale thermal behavior in a wide range of materials and devices.\cite{pollock_micro-thermal_2001,jeong_scanning_2015,gomes_scanning_2015,zhang_review_2020,bodzenta_scanning_2022}.  In this form of scanning probe microscopy (SPM), a sharp, thermally sensitive probe is scanned across a surface and the thermal signal is collected to create an image.  Several different types of thermal probe have been successfully implemented in SThM, relying on various physical phenomena such as the thermocouple effect\cite{majumdar_thermal_1995,kim_ultra-high_2012,shekhawat_micromachined_2018}, thermoresistance\cite{edinger_novel_2001,tovee_nanoscale_2012,menges_temperature_2016}, thermal-mechanical effects\cite{kim_thermal_2009}, or thermally modulated optical fluorescence\cite{aigouy_note_2011,tovee_mapping_2013}. Generally speaking, spatial resolutions of a few tens of nanometers have been achieved, allowing studies on many different nanoscale physical systems.

One of the most important considerations that has emerged in the SThM literature is the ambient environment in which the scanning probe is operated.  While a few studies have been conducted in vacuum\cite{hinz_high_2008,kim_ultra-high_2012,meng_temperature_2023} or liquid\cite{shi_thermal_2001,aigouy_note_2011,tovee_mapping_2013}, for the most part SThM has been conducted in air.  The presence of atmospheric-pressure air around the tip can lead to significant effects on the thermal signal, and these have led to much discussion in the literature.  Currently thinking postulates that if SThM is conducted in normal ambient laboratory atmosphere ($10^5$ Pa, $\approx50$ \% relative humidity), a water meniscus will form around the tip, and heat will 
\begin{figure}
\centering
\includegraphics[width=0.8\linewidth]{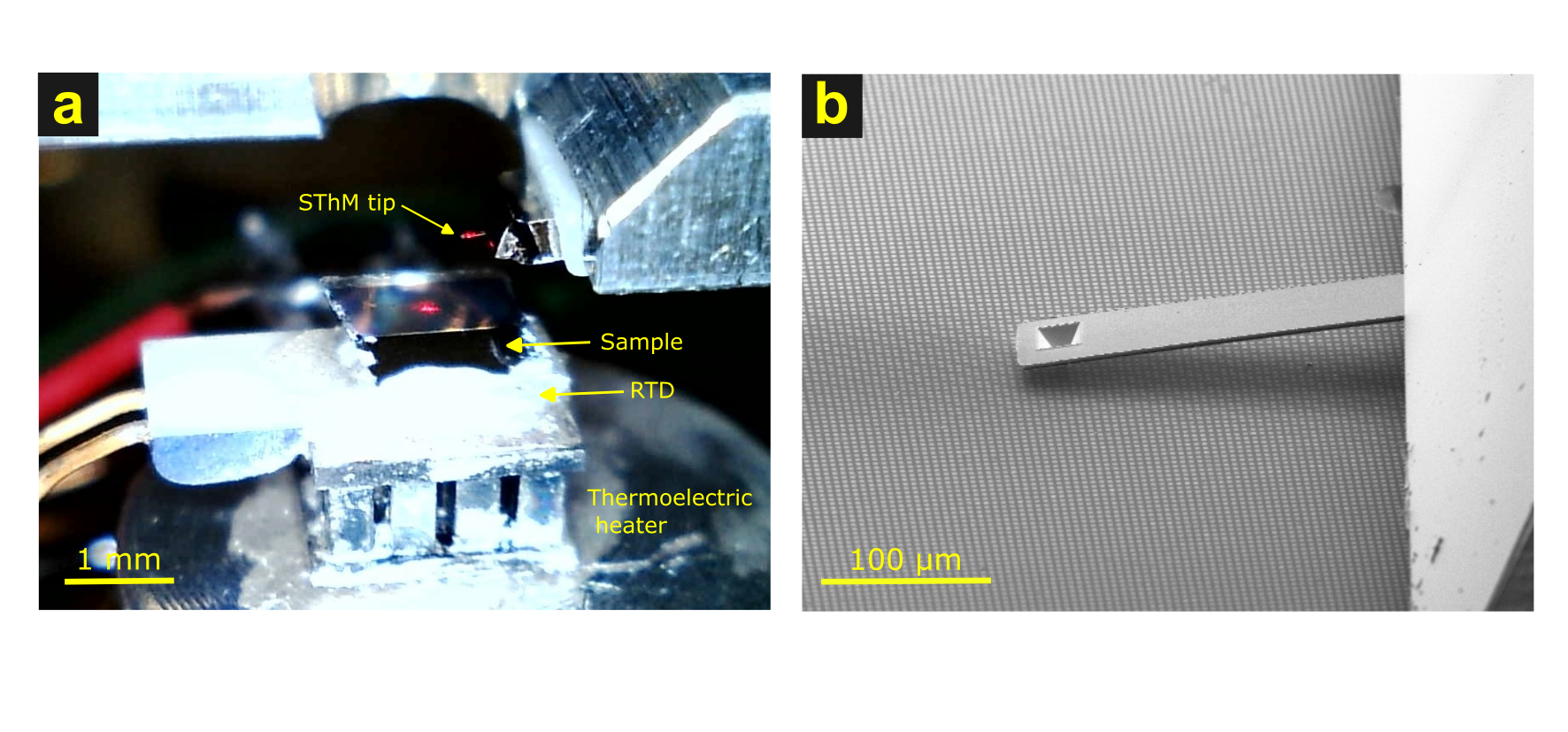}
\caption{(a) Photograph of the SThM tip and sample mounted on the resistance temperature detector (RTD) and thermoelectric heater. (b) SEM of the SThM tip landed on the sample.}
\label{fig:Figure01_SThMphoto_SEM}
\end{figure}
transfer through the surrounding air via conduction and convection from the surface to the sides of the tip. The water meniscus can lead to good thermal contact between the tip and the surface, boosting the measured thermal signal, but it also 
may have deleterious effects on the resolution, depending on its spatial extent.  The conductive and convective thermal transfer can result in a uniform background thermal signal in the most benign scenario, but 
may also result is spurious signals and resolution deterioration, depending on the nature of the sample being investigated.  Typically, if a SThM measurement is to be properly interpreted, the effects of the water meniscus and conduction/convection must be taken into consideration (see, for example, \cite{deshmukh_direct_2022}).

In this study, we conduct a direct comparison of SThM measurements in air and vacuum in order to elucidate some of the effects that ambient atmospheric conditions can have.  The experimental apparatus consists of a commercial SPM modified to perform SThM and integrated into the stage of a scanning electron microscope (SEM).  With this arrangement, it is possible to directly compare measurements in vacuum and air using the same thermal tip and same sample by either operating under normal SEM vacuum conditions, or venting the SEM and opening the door.

The SThM tip used in this work was of the thermocouple type\cite{shekhawat_micromachined_2018}. The sample was a silicon substrate with a nanofabricated array of 2.5 $\mu$m Ag squares encapsulated in SiO$_{2}$, where the surface was flattened by chemical-mechanical polishing, leaving a thin, 58 nm layer of SiO$_{2}$ over the Ag squares. The result was a sample with minimal surface topography, minimizing possible influence of topography on the thermal signal, and well-defined squares of high thermal conductivity to the surface.  When the sample was heated from below, the difference in thermal transport between the Ag squares and the surrounding SiO$_{2}$ was easily measured with the SThM. Experiments were conducted in steady state, with constant temperature applied to the bottom of the sample via a thermoelectric heater monitored and controlled by a temperature sensor; see Fig. \ref{fig:Figure01_SThMphoto_SEM}. The thermocouple voltage from the tip was amplified and recorded while the tip was scanned. Resolution was probed by examining line scans over the edge of the Ag squares.  

\section*{Results}
\noindent  In order to avoid possible artifacts due to tip-to-tip variations, all measurements presented here were done with a single SThM tip. The sample was heated to 101 $^{\circ}$C and this temperature was held constant via a servo loop.  Variations in this temperature, both drift and fluctuations, were less than 1 $^{\circ}$C. 

\begin{figure}[ht]
\centering
\includegraphics[width=0.8\linewidth]{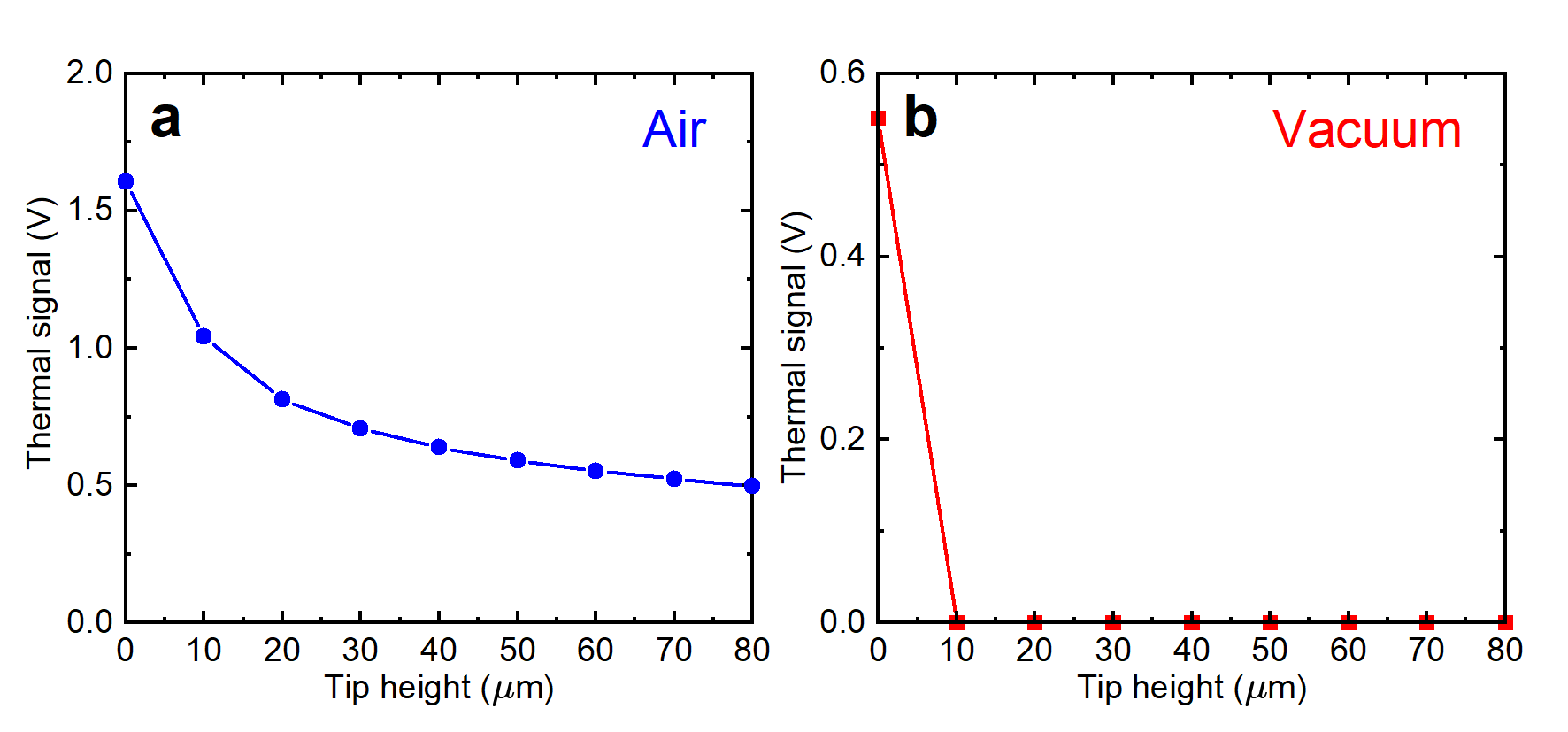}
\caption{Thermal signal as a function of tip height above the sample. (a) Air measurement, showing a drop of 37 \% when the tip leaves the sample, followed by a slow decline as the tip moves farther away. (b) Vacuum measurement, showing an abrupt drop at the first lift step when the tip loses contact with the sample. The thermal signal is the thermocouple voltage amplified by a factor of 2000 in the case of air, or 10,000 in the case of vacuum. Uncertainties are smaller than the symbols in these figures.}
\label{fig:Figure02_Sig_vs_tip_height}
\end{figure}

Figure \ref{fig:Figure02_Sig_vs_tip_height} shows the thermal signal as a function of tip lift height above the sample.  Starting at contact, the tip was raised in 10~$\mu$m increments while the signal was recorded. As seen in the figure, there is a marked difference between air and vacuum.  In air, the thermal signal drops by 37~\% when the tip is first lifted, and then only declines slowly, still maintaining 30 \% of the in-contact signal at a distance of 80~$\mu$m.  By contrast, the vacuum signal drops immediately to zero on lifting the tip, and remains there for all heights measured.  This is a clear indication of the effects of thermal transport to the tip through the air.  The repercussions of this effect depend strongly on the particular sample being measured.  In a simple case when the sample is flat and essentially uniformly heated, this effect can be expected to only contribute a flat background to a measurement, which can be easily subtracted.  On the other hand, if the sample has nearby local hot spots, or significant topography with varying temperatures, a measurement could be significantly distorted.  A vacuum measurement, on the other hand, contains no such artifacts and one can be assured that the signal comes only from direct contact between the tip and the sample. 

The thermal signal observed when the tip is in contact with the surface is of primary interest when conducting a SThM measurement. In an ideal world, the magnitude of this signal would be a measure of the surface temperature in the region contacted by the tip.  However for this to be true, the tip would have to represent a negligible thermal load on the contact region. In reality, the thermal load imposed by the tip is usually not negligible, and the thermal signal becomes dependent on the relative balance between the thermal properties of the tip itself, the thermal conductivity across the tip-surface junction, the thermal properties of the contact region, and the amount of heat flowing into the contact region.  In an air measurement, the relative balance of heat flowing out of the contact region via air conduction/convection vs heat flowing into the tip also plays a role.  If the air loss is significantly larger than the loss through the tip, then an assumption of negligible tip influence may actually be more applicable.  But if not, a more complex analysis must be carried out. In vacuum, heat flow out of the surface into free space can be assumed to be negligible (provided temperatures are not high enough to create significant radiative heat transfer), simplifying the analysis.  Heat flow into the tip across the tip-sample interface must still be considered, however, and can play an even more significant role in situations where air loss would otherwise dominate.  Ultimately, it is not generally possible to obtain a meaningful numerical value for the surface temperature; however, a relative measure of spatial variations in the thermal signal is often sufficient to provide insight into the nanoscale thermal properties of many systems. 
\begin{figure}[ht]
\centering
\includegraphics[width=0.9\linewidth]{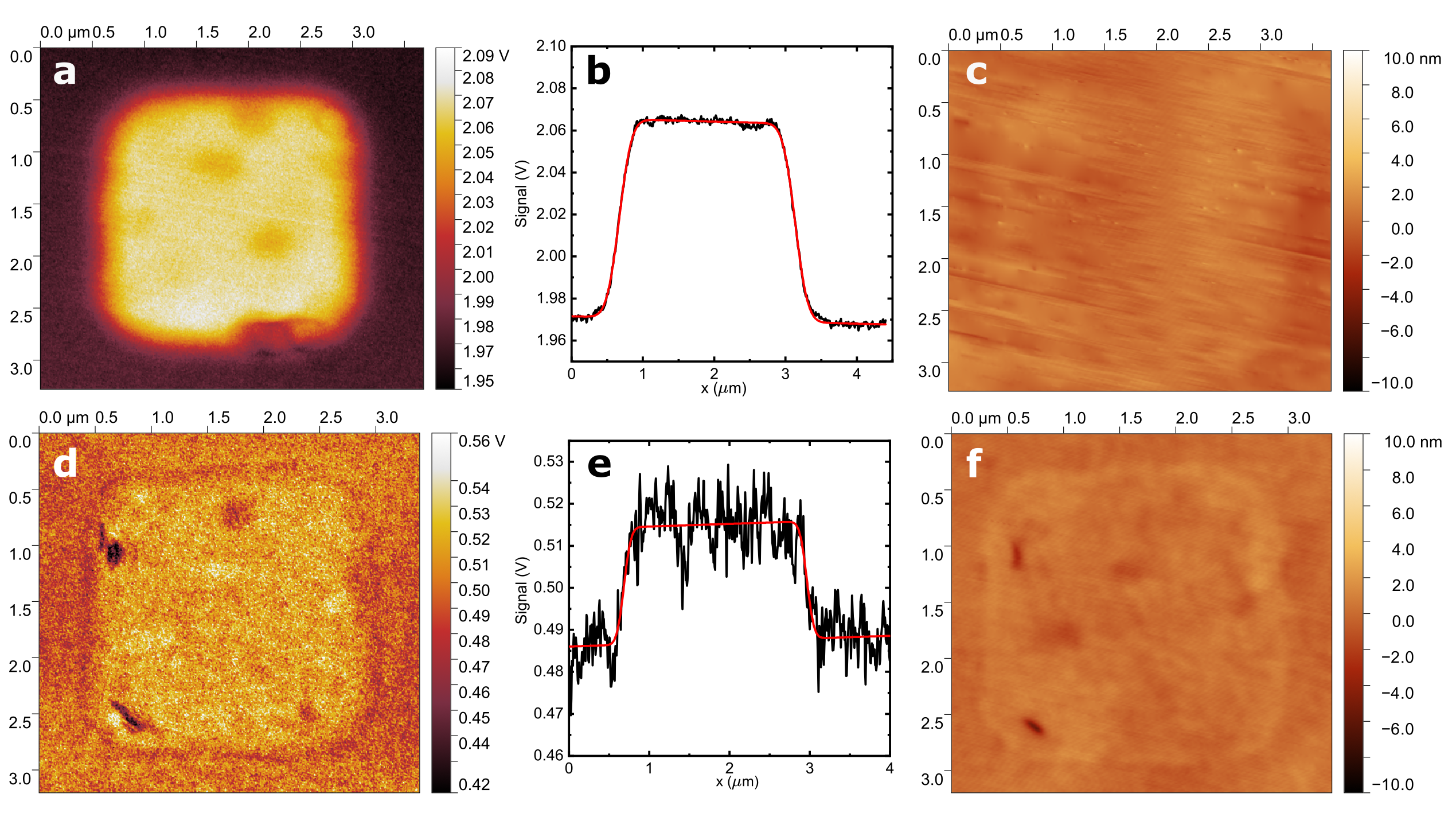}
\caption{(a) Thermal image of region over 2.5 $\mu$m Ag square recorded in air. (b) Horizontal line scan across image in (a). (c) Topography signal corresponding to (a). (d) Thermal image of region over 2.5 $\mu$m Ag square recorded in vacuum. (e) Horizontal line scan across image in (d). (f) Topography signal corresponding to (d). Thermal signal is thermocouple voltage after amplification by a factor of 2000 (air) or 10,000 (vacuum). Red line in (b) and (d) is fit to data as described in the text.}
\label{fig:Figure03_ThermalImage_linescan}
\end{figure}

A comparison of the magnitudes of the thermal signals seen in air and vacuum is also useful for shedding some light on the role played by an air environment in SThM.  For the specific tip used in the present experiments on the SiO$_{2}$ surface of the sample, we found the air signal to be generally quite reproducible and consistent.  Across nine landings in different locations on the sample held at 101 $^{\circ}$C, the average thermocouple voltage was (0.97$\pm$0.04) mV (uncertainty is one standard deviation about the mean).  On the other hand, the signal in vacuum was significantly smaller than in air, and was observed to have values that varied over a large range.  Thermocouple signals as small as 0.025 mV and as large as 0.37 mV were seen, depending on the recent history of the sample and tip.

On investigating the cause of the large variation in signal for the vacuum measurement, we found that the largest signals were seen when a measurement was done immediately after pumping down the SEM chamber following an air measurement.  The signal would typically remain high while scanning for about one to two hours, after which a rapid drop-off in signal was observed.  Once the signal dropped by a factor of about 5 to 10, it stabilized and remained at the low level essentially indefinitely.  Interestingly, the signal did not begin dropping until scanning commenced, suggesting that the scanning process was instigating the change.  Furthermore, once the signal had dropped, moving to a new location resulted in no change in signal, suggesting whatever change was happening was on the tip, not the sample. Also, re-exposing the tip or the sample to air via the SEM sample-exchange interlock did not cause any significant change in the signal once it had decayed.  The only thing that restored the vacuum signal to a high value was a scanned measurement in air. 

Figure \ref{fig:Figure03_ThermalImage_linescan} shows examples of SThM images collected in air and vacuum.  The difference in signal magnitude is clearly seen in the figure.  To investigate any difference in resolution, line scans across the images were fit to a function of the following form
\begin{equation}
    f(x) = -\frac{A}{2}\left(\erf\left[-\frac{x-x_1}{\sigma}\right] + \erf\left[\frac{x-x_2}{\sigma}\right]\right) + y_0 + sx, 
\end{equation}
where $A$, $x_1$, $\sigma$, $x_2$, $y_0$ and $s$ are free parameters.  The mathematical form for this function was chosen empirically to provide a reasonable fit. For the air measurements, the 25 \% to 75 \% edge width was extracted from the fits for nine images taken from different Ag squares. The average of these widths was found to be (180$\pm$18) nm.  For the vacuum measurements, similarly obtained widths from ten images were averaged to yield (139$\pm$36) nm (uncertainties are one standard deviation about the mean).  The result of these measurements is that the edge width appears to be somewhat smaller in vacuum than in air.  This difference becomes more significant if we take into account horizontal heat spreading in the 58 nm thickness of SiO$_{2}$ above the Ag square's edge.  We have estimated this effect by conducting a finite-element thermal simulation of a 20 nm rod moving across an Ag edge encapsulated in SiO$_{2}$.  The resulting simulated 25 \% to 75 \% edge width is 72 nm.  If one assumes a simple convolution of this with the SThM tip signal, it is reasonable to subtract it in quadrature from the measured values.  In this case the air width would be 165 nm and the vacuum width would be 119 nm, a difference of 39 \%.         

\section*{Discussion}
\noindent Our results show that, at least under the conditions described here, the thermal signal in air is consistently larger than it is in vacuum, and the resolution is somewhat lower (i.e., measured edge widths are somewhat larger). Furthermore, the thermal signal in vacuum can be highly variable, depending on the history of scanning while exposed to air.  These observations are entirely consistent with the formation of a water-related meniscus at the tip-surface interface when measurements are conducted in air. The liquid in the interface provides good thermal contact, increasing the heat flow into the tip, resulting in a higher thermocouple temperature in steady state. It also spreads the thermal contact out somewhat, reducing the resolution. The meniscus appears to be persistent during air measurements because the thermal signal remains stable. Interestingly, the meniscus apparently persists for some time after the SEM chamber is evacuated and the substrate is heated to over 100 $^\circ$C.  This persistent meniscus seems to be associated with the tip rather than the substrate, and its elimination appears to require scanning of the tip on the surface. After the meniscus eventually dissipates during a vacuum measurement, the thermal signal is essentially the same at all locations across the substrate.  If the drop in signal were associated with a change in the sample, one would expect the signal to still be larger is regions that had not been scanned yet. 

The persistence of the higher thermal signal at 100 °C in vacuum for an hour or more of scanning hints that the behavior may involve more than just a simple water meniscus; for example, a non-aqueous liquid at the tip-surface junction could be forming. According to Ievlev et al.,\cite{ievlev_chemical_2018}  commercial SPM probes that are stored in plastic boxes lined with sticky polydimethylsiloxane gel get contaminated with silicone oil. Upon contact with the sample surface, this oil starts flowing down the tip cone and spreading over the sample surface.  Depending on the amount of contamination, the oil may either form an oily meniscus or cover the water meniscus with an evaporation-protective film.  It could also chemically react with water (hydrolysis)\cite{garbay_measurements_1984} at elevated temperatures. These processes will stabilize the meniscus and hamper its evaporation or dissipation until the liquid is mechanically removed by rubbing the tip against the surface similar to dip-pen nanolithography\cite{piner_dip-pen_1999}.  These considerations also explain the apparent association of the meniscus with the tip, rather than the substrate, since it is the probe that is contaminated with the oil, not the sample. Re-exposure of the “depleted” probe to air may moisturize it again, increase the remnant oil mobility and facilitate the delivery of oil-water mixture to the tip apex.

To corroborate the hypothesis that a water-related meniscus plays a role in the vacuum signal variability, we conducted a further experiment where the sample was removed through the SEM interlock and wiped with a lens tissue saturated with purified water.  On returning the sample to the chamber and imaging with the SEM, clear patches were observed on the sample surface, presumably due to residual water.  Scanning the tip in these regions had the effect of temporarily restoring the thermal signal to the high level seen after an air measurement, even when scanning was done on areas that did not exhibit patches.  As with the other post-air vacuum measurements, after scanning for some time the signal returned to its lower level.  

The observation of a 39 \% larger edge width for the air measurements provides some insight into the influence that the water meniscus and conduction/convection can have on the resolution of an SThM measurement.  Of course, the extent of the effect is highly dependent on tip type and specific geometry, so it is difficult to generalize to other types of SThM.  Nevertheless, at least for the type of tip used in the present study, it appears the effect on resolution is measurable, but not too drastic. 

\begin{figure}
\centering
\includegraphics[width=0.8\linewidth]{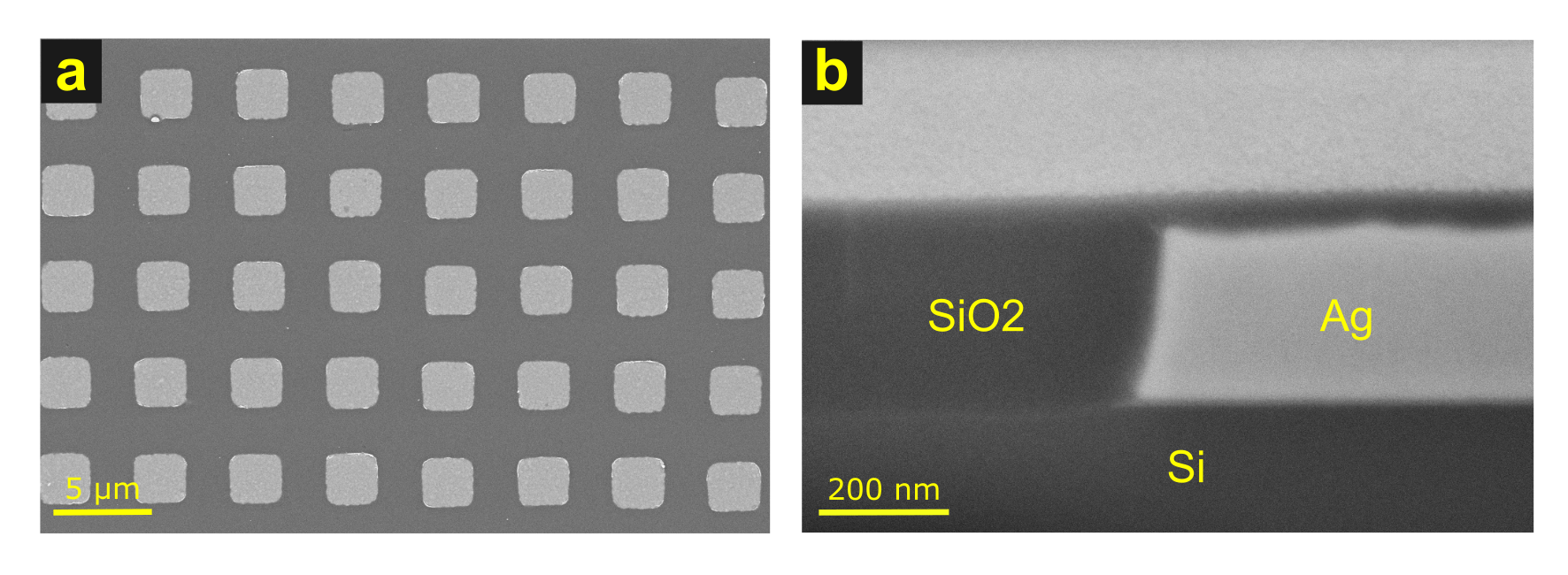}
\caption{SEM images of sample. (a) Top down view, showing Ag squares. The SEM beam energy was 5 keV, enabling imaging of the Ag through the thin SiO$_{2}$ covering. (b) Focused ion beam-milled cross section of edge of Ag square, viewed at an angle of 54$^{\circ}$.}
\label{fig:Figure04_Sample_SEMs}
\end{figure}

\section*{Methods}
\noindent The experiments described here were conducted using a multimode SPM mounted on the stage of a field-emission scanning electron microscope.  Typical vacuum levels inside the SEM were of order $10^{-4}$ Pa ($10^{-6}$ mbar) or less. The SPM was of the type where a laser beam is reflected from the cantilever and detected on a quadrant photodiode. Imaging was carried out in the conventional manner, with feedback controlled by the topography signal while the SThM thermocouple signal was recorded.  Collected images were calibrated and corrected for scanner skew by affine transformation, using the SEM images of the Ag squares as a reference. Both the tip and sample were interchangeable through the SEM sample-change interlock.  The tip holder was modified to accommodate thermocouple-type SThM tips, providing contacts for the thermocouple. An \textit{in situ} preamplifier mounted on the SEM stage provided a fixed gain of 100 before the thermal signal was routed through feedthroughs in the SEM door. This preamplified signal was further amplified by a factor of 20 to 100 using a low-noise voltage amplifier with a bandwidth of 3 kHz before it was fed into the external input of the SPM controller.  The bandwidth was chosen to be as low as possible while still ensuring negligible contribution to the SPM resolution at the chosen scan rate of 10 $\mu$m/s. 

The sample was mounted on a stack consisting of a thermoelectric cooler operated with reverse current to act as a heater and a platinum thin-film resistance temperature detector (RTD).  This stack was assembled using silver paint of the type typically used to mount SEM samples. The temperature of the stack was monitored by connecting the RTD to a Wheatstone bridge circuit.  The output of this circuit was passed through a 48 Hz low-pass filter, then fed into a PID (proportional-integral-differential) controller which governed the current flowing through the thermoelectric heater.  The set point of this controller was chosen to maintain a temperature of 101 $^{\circ}$C at the RTD, and this was observed to fluctuate less than 1 $^{\circ}$C when operated continuously in air or vacuum.  A finite-element thermal analysis showed that the sample surface was within (1 to 2) $^{\circ}$C of the temperature measured by the RTD.     

Fabrication of the sample began with a 0.5 mm thick p-doped Si wafer. Electron beam deposition was used to deposit a 10 nm Cr adhesion layer followed by 280 nm of Ag.  The Ag film was patterned into 2.5 $\mu$m squares on a 5 $\mu$m grid by laser lithography and ion milling.  The surface was then coated with $\approx$650 nm of SiO$_{2}$ using plasma-enhanced chemical vapor deposition. The final step involved chemical mechanical polishing to create a flat surface with a thin layer of SiO$_{2}$ over the Ag squares.  After polishing, the root-mean-square roughness of the surface was less than 2 nm in several sampled areas across the sample, with the exception of occasional small defects near the corners of the Ag squares. Figure \ref{fig:Figure04_Sample_SEMs} shows top-down and cross sectional views of the resulting embedded Ag structures. The cross-sectional image was used to measure the Ag and SiO$_{2}$ thicknesses to be (58$\pm$3) nm and (280$\pm$14) nm, respectively, where the uncertainties are combined one standard deviation arising from surface roughness, edge identification uncertainty, and SEM calibration.  

\begin{figure}
\centering
\includegraphics[width=0.7\linewidth]{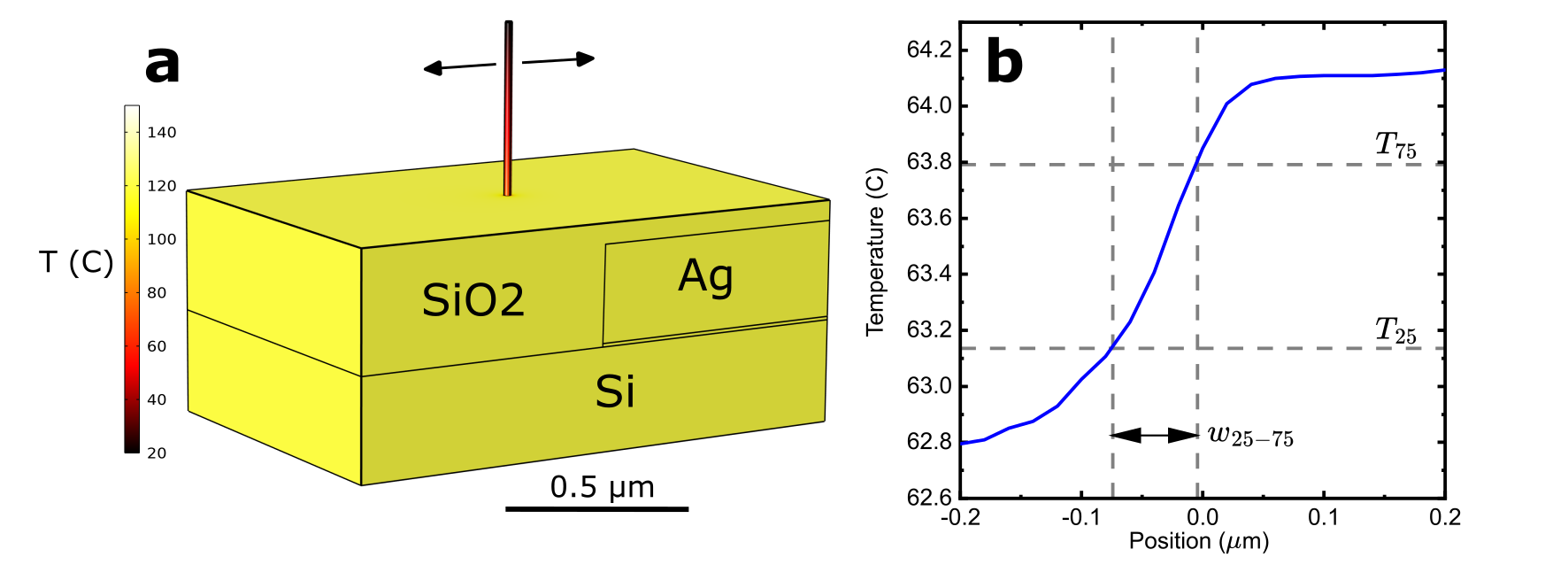}
\caption{Finite-element calculation of signal spread due to thermal diffusion between Ag square edge and surface of SiO$_{2}$. (a) Geometry of calculation, showing Si substrate, Ag square, SiO$_{2}$ coating, and Cr rod used to represent the tip. (b) Plot of temperature measured at a point inside the rod a distance 50 nm above the surface as a function of rod position. Dashed lines indicate the 25 \% and 75 \% temperature values $T_{25}$ and $T_{75}$, respectively, and their corresponding positions. The 25-75 width $w_{25-75}$ = 72 nm is indicated in the figure.}
\label{fig:Figure05_COMSOL}
\end{figure}

To estimate the horizontal thermal spreading between the Ag square edge and the SiO$_{2}$ surface, we used a commercial finite-element simulation package.  The SThM measurement was represented by scanning a 20 nm diameter Cr rod across the surface of a SiO$_{2}$ sample with buried Ag region (See Fig. \ref{fig:Figure05_COMSOL}).  The distal end of the rod was fixed at 20 $^{\circ}$C and its interior temperature at a point 50 nm above the surface was recorded.  The resulting temperature profile provides a reasonable approximation of what a thermocouple SThM tip would measure, where we have chosen the rod diameter to be small enough so that the profile measured represents an essentially unconvoluted measure of the thermal spread.

\section*{Conclusions}
We have carried out a direct comparison of SThM measurements conducted in air and vacuum and shown that while air measurements can have a higher signal level and be more stable and reproducible than vacuum measurements, they may suffer from somewhat poorer resolution.  We have seen direct evidence of the effects of the formation of a meniscus when samples are scanned in air, and observed that the associated high thermal conductivity can persist in vacuum for several hours of scanning before it dissipates.  The present study is of course limited in the type of tip used, the type of the sample, and the type of environments investigated.  Nevertheless, it provides useful input into the ongoing discussion in the literature of the role played by the meniscus and conduction/convection effects when SThM is conducted in ambient air.  

The present studies do not allow differentiation between conduction/convection or meniscus effects on the resolution.  To differentiate between these effects it would be interesting to perform measurements in dry air or other inert atmospheres, where conduction/convection would still be present but a water meniscus presumably would not.  Such studies will need to be carried out in the future in order to provide a full picture of the complex effects of ambient environments on SThM measurements.      

\clearpage
\bibliography{references}

\begin{thebibliography}{10}
\urlstyle{rm}
\expandafter\ifx\csname url\endcsname\relax
  \def\url#1{\texttt{#1}}\fi
\expandafter\ifx\csname urlprefix\endcsname\relax\def\urlprefix{URL }\fi
\expandafter\ifx\csname doiprefix\endcsname\relax\def\doiprefix{DOI: }\fi
\providecommand{\bibinfo}[2]{#2}
\providecommand{\eprint}[2][]{\url{#2}}

\bibitem{pollock_micro-thermal_2001}
\bibinfo{author}{Pollock, H.~M.} \& \bibinfo{author}{Hammiche, A.}
\newblock \bibinfo{journal}{\bibinfo{title}{Micro-thermal analysis: techniques and applications}}.
\newblock {\emph{\JournalTitle{Journal of Physics D: Applied Physics}}} \textbf{\bibinfo{volume}{34}}, \bibinfo{pages}{R23}, \doiprefix\url{10.1088/0022-3727/34/9/201} (\bibinfo{year}{2001}).

\bibitem{jeong_scanning_2015}
\bibinfo{author}{Jeong, W.}, \bibinfo{author}{Hur, S.}, \bibinfo{author}{Meyhofer, E.} \& \bibinfo{author}{Reddy, P.}
\newblock \bibinfo{journal}{\bibinfo{title}{Scanning {Probe} {Microscopy} for {Thermal} {Transport} {Measurements}}}.
\newblock {\emph{\JournalTitle{Nanoscale and Microscale Thermophysical Engineering}}} \textbf{\bibinfo{volume}{19}}, \bibinfo{pages}{279--302}, \doiprefix\url{10.1080/15567265.2015.1109740} (\bibinfo{year}{2015}).
\newblock \bibinfo{note}{Place: Philadelphia Publisher: Taylor \& Francis Inc WOS:000366663600004}.

\bibitem{gomes_scanning_2015}
\bibinfo{author}{Gomès, S.}, \bibinfo{author}{Assy, A.} \& \bibinfo{author}{Chapuis, P.-O.}
\newblock \bibinfo{journal}{\bibinfo{title}{Scanning thermal microscopy: {A} review}}.
\newblock {\emph{\JournalTitle{physica status solidi (a)}}} \textbf{\bibinfo{volume}{212}}, \bibinfo{pages}{477--494}, \doiprefix\url{10.1002/pssa.201400360} (\bibinfo{year}{2015}).
\newblock \bibinfo{note}{\_eprint: https://onlinelibrary.wiley.com/doi/pdf/10.1002/pssa.201400360}.

\bibitem{zhang_review_2020}
\bibinfo{author}{Zhang, Y.} \emph{et~al.}
\newblock \bibinfo{journal}{\bibinfo{title}{A {Review} on {Principles} and {Applications} of {Scanning} {Thermal} {Microscopy} ({SThM})}}.
\newblock {\emph{\JournalTitle{Advanced Functional Materials}}} \textbf{\bibinfo{volume}{30}}, \bibinfo{pages}{1900892}, \doiprefix\url{10.1002/adfm.201900892} (\bibinfo{year}{2020}).
\newblock \bibinfo{note}{\_eprint: https://onlinelibrary.wiley.com/doi/pdf/10.1002/adfm.201900892}.

\bibitem{bodzenta_scanning_2022}
\bibinfo{author}{Bodzenta, J.} \& \bibinfo{author}{Kaźmierczak-Bałata, A.}
\newblock \bibinfo{journal}{\bibinfo{title}{Scanning thermal microscopy and its applications for quantitative thermal measurements}}.
\newblock {\emph{\JournalTitle{Journal of Applied Physics}}} \textbf{\bibinfo{volume}{132}}, \bibinfo{pages}{140902}, \doiprefix\url{10.1063/5.0091494} (\bibinfo{year}{2022}).

\bibitem{majumdar_thermal_1995}
\bibinfo{author}{Majumdar, A.} \emph{et~al.}
\newblock \bibinfo{journal}{\bibinfo{title}{Thermal imaging by atomic force microscopy using thermocouple cantilever probes}}.
\newblock {\emph{\JournalTitle{Review of Scientific Instruments}}} \textbf{\bibinfo{volume}{66}}, \bibinfo{pages}{3584--3592}, \doiprefix\url{10.1063/1.1145474} (\bibinfo{year}{1995}).

\bibitem{kim_ultra-high_2012}
\bibinfo{author}{Kim, K.}, \bibinfo{author}{Jeong, W.}, \bibinfo{author}{Lee, W.} \& \bibinfo{author}{Reddy, P.}
\newblock \bibinfo{journal}{\bibinfo{title}{Ultra-{High} {Vacuum} {Scanning} {Thermal} {Microscopy} for {Nanometer} {Resolution} {Quantitative} {Thermometry}}}.
\newblock {\emph{\JournalTitle{ACS Nano}}} \textbf{\bibinfo{volume}{6}}, \bibinfo{pages}{4248--4257}, \doiprefix\url{10.1021/nn300774n} (\bibinfo{year}{2012}).
\newblock \bibinfo{note}{Publisher: American Chemical Society}.

\bibitem{shekhawat_micromachined_2018}
\bibinfo{author}{Shekhawat, G.~S.} \emph{et~al.}
\newblock \bibinfo{journal}{\bibinfo{title}{Micromachined {Chip} {Scale} {Thermal} {Sensor} for {Thermal} {Imaging}}}.
\newblock {\emph{\JournalTitle{ACS Nano}}} \textbf{\bibinfo{volume}{12}}, \bibinfo{pages}{1760--1767}, \doiprefix\url{10.1021/acsnano.7b08504} (\bibinfo{year}{2018}).
\newblock \bibinfo{note}{Publisher: American Chemical Society}.

\bibitem{edinger_novel_2001}
\bibinfo{author}{Edinger, K.}, \bibinfo{author}{Gotszalk, T.} \& \bibinfo{author}{Rangelow, I.~W.}
\newblock \bibinfo{journal}{\bibinfo{title}{Novel high resolution scanning thermal probe}}.
\newblock {\emph{\JournalTitle{Journal of Vacuum Science \& Technology B: Microelectronics and Nanometer Structures Processing, Measurement, and Phenomena}}} \textbf{\bibinfo{volume}{19}}, \bibinfo{pages}{2856--2860}, \doiprefix\url{10.1116/1.1420580} (\bibinfo{year}{2001}).

\bibitem{tovee_nanoscale_2012}
\bibinfo{author}{Tovee, P.}, \bibinfo{author}{Pumarol, M.}, \bibinfo{author}{Zeze, D.}, \bibinfo{author}{Kjoller, K.} \& \bibinfo{author}{Kolosov, O.}
\newblock \bibinfo{journal}{\bibinfo{title}{Nanoscale spatial resolution probes for scanning thermal microscopy of solid state materials}}.
\newblock {\emph{\JournalTitle{Journal of Applied Physics}}} \textbf{\bibinfo{volume}{112}}, \bibinfo{pages}{114317}, \doiprefix\url{10.1063/1.4767923} (\bibinfo{year}{2012}).

\bibitem{menges_temperature_2016}
\bibinfo{author}{Menges, F.} \emph{et~al.}
\newblock \bibinfo{journal}{\bibinfo{title}{Temperature mapping of operating nanoscale devices by scanning probe thermometry}}.
\newblock {\emph{\JournalTitle{Nature Communications}}} \textbf{\bibinfo{volume}{7}}, \bibinfo{pages}{10874}, \doiprefix\url{10.1038/ncomms10874} (\bibinfo{year}{2016}).
\newblock \bibinfo{note}{Publisher: Nature Publishing Group}.

\bibitem{kim_thermal_2009}
\bibinfo{author}{Kim, S.-J.}, \bibinfo{author}{Ono, T.} \& \bibinfo{author}{Esashi, M.}
\newblock \bibinfo{journal}{\bibinfo{title}{Thermal imaging with tapping mode using a bimetal oscillator formed at the end of a cantilever}}.
\newblock {\emph{\JournalTitle{Review of Scientific Instruments}}} \textbf{\bibinfo{volume}{80}}, \bibinfo{pages}{033703}, \doiprefix\url{10.1063/1.3095680} (\bibinfo{year}{2009}).

\bibitem{aigouy_note_2011}
\bibinfo{author}{Aigouy, L.}, \bibinfo{author}{Lalouat, L.}, \bibinfo{author}{Mortier, M.}, \bibinfo{author}{Löw, P.} \& \bibinfo{author}{Bergaud, C.}
\newblock \bibinfo{journal}{\bibinfo{title}{Note: {A} scanning thermal probe microscope that operates in liquids}}.
\newblock {\emph{\JournalTitle{Review of Scientific Instruments}}} \textbf{\bibinfo{volume}{82}}, \bibinfo{pages}{036106}, \doiprefix\url{10.1063/1.3567794} (\bibinfo{year}{2011}).

\bibitem{tovee_mapping_2013}
\bibinfo{author}{Tovee, P.~D.} \& \bibinfo{author}{Kolosov, O.~V.}
\newblock \bibinfo{journal}{\bibinfo{title}{Mapping nanoscale thermal transfer in-liquid environment—immersion scanning thermal microscopy}}.
\newblock {\emph{\JournalTitle{Nanotechnology}}} \textbf{\bibinfo{volume}{24}}, \bibinfo{pages}{465706}, \doiprefix\url{10.1088/0957-4484/24/46/465706} (\bibinfo{year}{2013}).
\newblock \bibinfo{note}{Publisher: IOP Publishing}.

\bibitem{hinz_high_2008}
\bibinfo{author}{Hinz, M.}, \bibinfo{author}{Marti, O.}, \bibinfo{author}{Gotsmann, B.}, \bibinfo{author}{Lantz, M.~A.} \& \bibinfo{author}{Duerig, U.}
\newblock \bibinfo{journal}{\bibinfo{title}{High resolution vacuum scanning thermal microscopy of {HfO2} and {SiO2}}}.
\newblock {\emph{\JournalTitle{Applied Physics Letters}}} \textbf{\bibinfo{volume}{92}}, \bibinfo{pages}{043122}, \doiprefix\url{10.1063/1.2840186} (\bibinfo{year}{2008}).
\newblock \bibinfo{note}{Place: Melville Publisher: Amer Inst Physics WOS:000252860400097}.

\bibitem{meng_temperature_2023}
\bibinfo{author}{Meng, J.} \emph{et~al.}
\newblock \bibinfo{journal}{\bibinfo{title}{Temperature {Distribution} in {TaOx} {Resistive} {Switching} {Devices} {Assessed} {In} {Operando} by {Scanning} {Thermal} {Microscopy}}}.
\newblock {\emph{\JournalTitle{ACS Applied Electronic Materials}}} \textbf{\bibinfo{volume}{5}}, \bibinfo{pages}{2414--2421}, \doiprefix\url{10.1021/acsaelm.3c00229} (\bibinfo{year}{2023}).
\newblock \bibinfo{note}{Publisher: American Chemical Society}.

\bibitem{shi_thermal_2001}
\bibinfo{author}{Shi, L.} \& \bibinfo{author}{Majumdar, A.}
\newblock \bibinfo{journal}{\bibinfo{title}{Thermal {Transport} {Mechanisms} at {Nanoscale} {Point} {Contacts}}}.
\newblock {\emph{\JournalTitle{Journal of Heat Transfer}}} \textbf{\bibinfo{volume}{124}}, \bibinfo{pages}{329--337}, \doiprefix\url{10.1115/1.1447939} (\bibinfo{year}{2001}).

\bibitem{deshmukh_direct_2022}
\bibinfo{author}{Deshmukh, S.} \emph{et~al.}
\newblock \bibinfo{journal}{\bibinfo{title}{Direct measurement of nanoscale filamentary hot spots in resistive memory devices}}.
\newblock {\emph{\JournalTitle{Science Advances}}} \textbf{\bibinfo{volume}{8}}, \bibinfo{pages}{eabk1514}, \doiprefix\url{10.1126/sciadv.abk1514} (\bibinfo{year}{2022}).
\newblock \bibinfo{note}{Publisher: American Association for the Advancement of Science}.

\bibitem{ievlev_chemical_2018}
\bibinfo{author}{Ievlev, A.~V.} \emph{et~al.}
\newblock \bibinfo{journal}{\bibinfo{title}{Chemical {Phenomena} of {Atomic} {Force} {Microscopy} {Scanning}}}.
\newblock {\emph{\JournalTitle{Analytical Chemistry}}} \textbf{\bibinfo{volume}{90}}, \bibinfo{pages}{3475--3481}, \doiprefix\url{10.1021/acs.analchem.7b05225} (\bibinfo{year}{2018}).
\newblock \bibinfo{note}{Publisher: American Chemical Society}.

\bibitem{garbay_measurements_1984}
\bibinfo{author}{Garbay, H.}, \bibinfo{author}{Grob, R.}, \bibinfo{author}{Casanovas, J.} \& \bibinfo{author}{Crine, J.-P.}
\newblock \bibinfo{title}{Measurements and influence of water content in silicone oil}.
\newblock In \emph{\bibinfo{booktitle}{1984 {IEEE} {International} {Conference} on {Eletrical} {Insulation}}}, \bibinfo{pages}{297--300}, \doiprefix\url{10.1109/EIC.1984.7465202} (\bibinfo{year}{1984}).

\bibitem{piner_dip-pen_1999}
\bibinfo{author}{Piner, R.~D.}, \bibinfo{author}{Zhu, J.}, \bibinfo{author}{Xu, F.}, \bibinfo{author}{Hong, S.} \& \bibinfo{author}{Mirkin, C.~A.}
\newblock \bibinfo{journal}{\bibinfo{title}{"{Dip}-{Pen}" {Nanolithography}}}.
\newblock {\emph{\JournalTitle{Science}}} \textbf{\bibinfo{volume}{283}}, \bibinfo{pages}{661--663}, \doiprefix\url{10.1126/science.283.5402.661} (\bibinfo{year}{1999}).
\newblock \bibinfo{note}{Publisher: American Association for the Advancement of Science}.

\end{thebibliography}

\section*{Acknowledgments}
The authors would like to thank Daron Westly And Richard Kasica for useful discussions concerning sample preparation.

\section*{Author contributions statement}

J. M. conceived the experiments, analyzed the results and prepared the manuscript. E. S. assisted with the apparatus and contributed to the manuscript.  A. C. provided intellectual input and information on SThM tips.  All authors reviewed the manuscript. 

\section*{Additional information}
The authors declare no competing interests.

\end{document}